\documentclass[aps,prl,twocolumn,superscriptaddress,floatfix]{revtex4}
\usepackage{graphicx,epsfig,amsmath}

\newcommand{\subfig}[2]{Fig.~\ref{fig:#1}(#2)} %for instance \subfig{LargeDiamonds}{b}

\newcommand{\Vone}{\mbox{$\text{V}_{\text{1}}$}}
\newcommand{\Vtwo}{\mbox{$\text{V}_{\text{2}}$}}

\newcommand{\NL}{\mbox{$\text{N}_{\text{L}}$}}
\newcommand{\NC}{\mbox{$\text{N}_{\text{C}}$}}
\newcommand{\NR}{\mbox{$\text{N}_{\text{R}}$}}
\newcommand{\Tts}{\mbox{$\text{T}_{\text{2}}^{\text{*}}$}}
\newcommand{\Tminus}{\mbox{$\text{T}_{\text{-}}$}}
\newcommand{\Tzero}{\mbox{$\text{T}_{\text{0}}$}}
\newcommand{\Tplus}{\mbox{$\text{T}_{\text{+}}$}}

\begin{document}

% The title
\title{Quantum interference between three two-spin states  in a double quantum dot }

% The authors

\author{S.~A.~Studenikin}
	\affiliation{Institute for Microstructural Sciences, National Research Council Canada, Ottawa, ON Canada K1A 0R6}
\author{G.~C.~Aers}
	\affiliation{Institute for Microstructural Sciences, National Research Council Canada, Ottawa, ON Canada K1A 0R6}
\author{G.~Granger}
		\affiliation{Institute for Microstructural Sciences, National Research Council Canada, Ottawa, ON Canada K1A 0R6}
\author{L.~Gaudreau}
	\affiliation{Institute for Microstructural Sciences, National Research Council Canada, Ottawa, ON Canada K1A 0R6}
	\affiliation{D\'epartement de physique, Universit\'e de Sherbrooke, Sherbrooke, QC Canada J1K 2R1}
\author{A.~Kam}
	\affiliation{Institute for Microstructural Sciences, National Research Council Canada, Ottawa, ON Canada K1A 0R6}
\author{P.~Zawadzki}
	\affiliation{Institute for Microstructural Sciences, National Research Council Canada, Ottawa, ON Canada K1A 0R6}
\author{Z.~R.~Wasilewski}
	\affiliation{Institute for Microstructural Sciences, National Research Council Canada, Ottawa, ON Canada K1A 0R6}
\author{A.~S.~Sachrajda}
  \email{Andrew.Sachrajda@nrc.ca}
	\affiliation{Institute for Microstructural Sciences, National Research Council Canada, Ottawa, ON Canada K1A 0R6}

% date
%\date{\today}

% The abstract
\begin{abstract}

Qubits based on the singlet (S) and the triplet ($\Tzero$, $\Tplus$) states  in double quantum dots have been demonstrated
in separate  experiments.  It has been recently proposed theoretically that under certain conditions a quantum interference
could occur from the interplay between these two qubit species. Here we report experiments and modeling which confirm these
theoretical predictions and identify the conditions under which this interference occurs.  Density matrix calculations show
that the interference pattern manifests primarily via the occupation of the common singlet state.  The S/$\Tzero$ qubit is
found to have a much longer $\Tts$  as  compared to the S/$\Tplus$  qubit.

\end{abstract}

% The PACS:

% make the title
\maketitle

% begin the article:	

%\section{Introduction}

	Recently two semiconductor based qubits have been demonstrated individually in double quantum dots,
based on a singlet and two different triplet states (S/$\Tzero$ and S/$\Tplus$) of two interacting spins.
These qubits possess the advantageous property for qubit addressability in that quantum gate operations can
be achieved by purely electrostatic means. The singlet and triplet states differ in their spin and, therefore,
can interact (to form a qubit) due to  the small statistical magnetic field gradients between the dots originating
from the nuclear spins and the hyperfine interaction.  As more complex quantum circuits and operations are
developed \cite{Gaudreau2011, Gaudreau2009, Granger2010} this raises the question whether and how these two qubits
would interfere since they both include the singlet as a component state.  In experiments to date this question
has been purposely avoided  by  passing through the S/$\Tplus$ anticrossing fast enough to avoid involving the
$\Tplus$ state \cite{Petta2005} or by keeping far enough away from the $S/\Tzero$  interaction region.\cite{Petta2010}
 However, in a recent theoretical paper on the coherent control of a two-electron spin system \cite{Sarkka2011}
S\"arkk\"a and Harju predicted that by using suitable pulses the two qubits should coexist resulting in a more
complex pattern of coherent behavior with all three states involved. Here we present experimental and theoretical
results confirming this  prediction of an interplay between the singlet and triplet qubits  in a double quantum dot.

\begin{figure}
\includegraphics[width=1.0 \columnwidth]{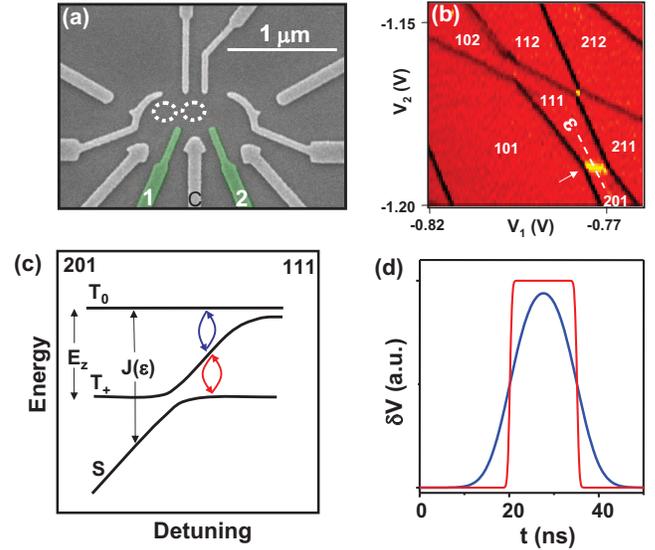} \vskip 0.0cm
\caption{(a) Electron micrograph of the  triple dot device. Fast voltage pulses ($\delta
V_1$,$\delta V_2$) are applied to gates 1 and 2 in addition to DC voltages (\Vone,\Vtwo). Gate
C tunes the (1,1,1) region size and can be adjusted such that the pair of dots indicated by the
white ovals operate independently of the right  ''spectator dot".  (b) Stability diagram
obtained from numerically differentiating the left QPC detector conductance with respect to
\Vtwo~at B=0.2~T. Black is low, red is medium, and yellow is high. Charge addition lines appear
black, and charge transfer lines appear yellow \cite{Granger2010}. A  detuning line is
drawn across the (2,0,1)/(1,1,1) charge transfer line indicated by the white arrow.  (c) Schematic energy diagram  of the
two-spin states near the (2,0,1)/(1,1,1) charge transfer line. (d) Examples of the pulse shapes for a
pulse duration $\tau$=15~ns after Gaussian filtering, leading to rise times of 8.0 (blue) and 0.8~ns (red).
 }
 \label{fig:1}
\end{figure}

For the experimental observation of the two kinds of singlet/triplet qubits we utilize a linear triple dot
device \cite{Gaudreau2009, Granger2010, Gaudreau2011} with gate voltages adjusted so that one of the dots acts as
a ''spectator"  with exchange energy to the central dot close to zero in the detuning range of interest. Under
these conditions the remaining  two dots may be regarded as a double dot and the relevant states can be described
in the language of singlet and triplet states \cite{Petta2005, Petta2010}.

An SEM image of the device is shown in \subfig{1}{a}.
For these experiments charge detection measurements are made with the quantum point contact (QPC) \cite{Field1993}
on the left side of the device.  The charge state of the device  and tunneling between dots  are controlled using
a combination of gates 1 and 2, which are also connected to high frequency lines.

The charge detection stability diagram obtained in the absence of pulses is shown in \subfig{1}{b}.  We focus on
coherent spin manipulation in a spin qubit regime  between the (\NL,\NC,\NR)=(1,1,1) and (2,0,1) electronic charge
configurations, or  (1,1)/(2,0) in the double dot notation. The C gate is tuned to make the (1,1,1) region wide
enough ($>$15 mV along \Vtwo) such that the right dot contribution is negligible.  The two dots acting as a
double dot are indicated by the white ovals in \subfig{1}{a}. The white dashed line in \subfig{1}{b} illustrates a
detuning line $\epsilon$ whose $\Vtwo$ component corresponds to the abscissa in \subfig{1}{c}.  In the two dot language used
in the remainder of this letter this detuning line crosses the (2,0)/(1,1) charge transfer line.

\begin{figure}
\includegraphics[width=1.0\columnwidth]{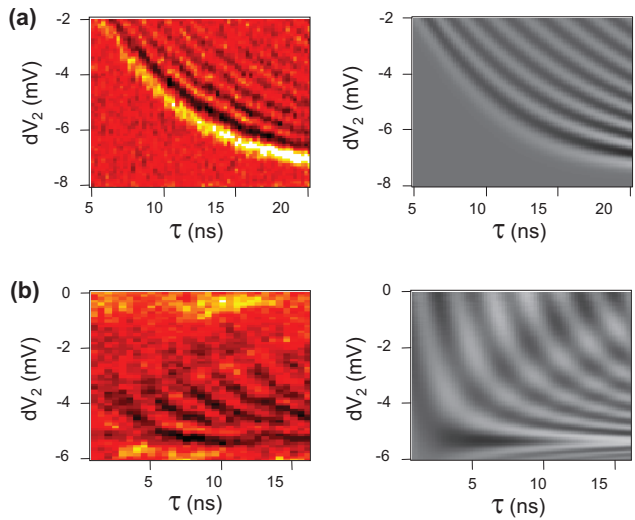}\vskip 0.0cm
\caption{(a,b) Left panels: experimental maps in the $\tau-\Vtwo$~plane
from the numerical derivative of the QPC conductance with respect to detuning component  $\Vtwo$ for the pulse rise times (a)
8.0 ns and (b) 0.8 ns.  For (a)[(b)] the pulse involves ($\delta V_1$,$\delta V_2$)=(-5,10)~mV [(-4,8)~mV] to traverse the
charge transfer line between (2,0) and (1,1) and repeats every $T_m$=0.5~$\mu$s, B=0.08~T. Black is low, orange is
medium, and yellow is high transconductance. Right panels: calculated derivative with respect to detuning of singlet
probability $P_{S}$ in the $\tau-\Vtwo$ plane for the rise times of the left panels.
Black (white) is low (high). $dV_2$ shift is due to 2 mV pulse change between (a) and (b).}
\label{fig:2}
\end{figure}

The lowest electronic states of a double dot containing two electrons consist of two  singlet states, S(2,0) and S(1,1),
and three triplet states, $\Tminus$, $\Tzero$, and  $\Tplus$. The latter are split by the Zeeman energy in an applied
in-plane magnetic  field. The two singlet states anticross as a function of detuning due to charge coupling between the dots.
By changing gate voltages to move along the detuning dashed line $\epsilon$ in \subfig{1}{b} the ground state singlet S can be tuned to
cross the $\Tplus$(1,1) triplet state and approach the $\Tzero$(1,1) state asymptotically
as illustrated in \subfig{1}{c}.\cite{Petta2005, Petta2010, Taylor2007} The S and T$_0$ states are split by the exchange
energy J($\epsilon$). The S and T$_+$ states cross at a detuning which depends upon magnetic field.  Nuclear hyperfine
field gradients between the dots cause the S/T$_+$ crossing to become an anticrossing and the S/T$_0$ spacing to be
asymptotically nonzero.  Here we plot the triplet states as horizontal lines as in \cite {Taylor2007} and restrict the
diagram to only three relevant levels S, T$_+$ and T$_0$.

 The spin dynamics in response to a voltage pulse is calculated from the time dependence of the density
matrix $\rho$ in the S/$\Tplus$/$\Tzero$ system with a Hamiltonian:

\begin{equation} \label{eqn:hamiltonian} H= \begin{pmatrix} E_{T_{0}} & 0 &
\Gamma_{S,{T_{0}}}\\ 0 & E_{T_{+}} & \Gamma_{S,{T_{+}}}\\  \Gamma_{S,{T_{0}}}^*
& \Gamma_{S,{T_{+}}}^* & E_{S} \end{pmatrix} \end{equation}
where the energies on the diagonal are those in the absence of nuclear
hyperfine  interactions. These hyperfine terms appear as $\Gamma_{S,{T_{+}}}$ and
$\Gamma_{S,{T_{0}}}$ which are respectively the differences between the dots of
the (x,y) and z hyperfine fields \cite{Taylor2007}. In all the calculations shown here
we take $\Gamma_{S,{T_{+}}}$ = $\Gamma_{S,{T_{0}}}$ = 0.2~$\mu$eV (a typical value to
fit the observed experimental fringe contrast\cite{Gaudreau2011}).

The time evolution of $\rho$ is calculated from an initial state at large
negative detuning where $P_{S}$=1, using:

\begin{equation} \label{eqn:rho} \frac{d\rho}{dt}=i\left[\rho , H/\hbar \right]
\end{equation}
This yields a set of three differential equations solved numerically by the Runge-Kutta method.

\begin{figure}
\includegraphics[width=1.0\columnwidth]{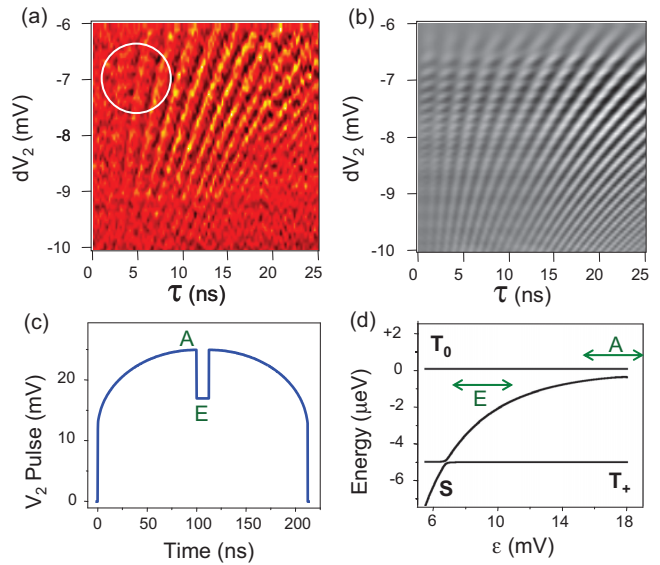} \vskip 0.0cm
\caption{(a): Experimental map in the $\tau-\Vtwo$~plane from the numerical derivative of the QPC conductance with respect to detuning component  $\Vtwo$ for an elliptical   pulse rise shown in (c), repetition time $T_m=2\mu s$,  B=0.2~T. Circle indicates region showing non-adiabatic structure on fringes. Black is low, orange is medium, and yellow is high transconductance. (b): Calculated derivative with respect to detuning of $P_{S}$ in the $\tau-\Vtwo$ plane. Black (white) is low (high). The origin for the pulse detuning is the charge transfer line. (c): Pulse shape added to the DC detuning. (d) Energy diagram, including hyperfine splitting, showing the detuning ranges of the adiabatic (A) and exchange (E) steps. The effective energy/gate voltage lever arm used here is 38.1 $\mu$eV/mV.}
\label{fig:3} \end{figure}

Filtering a rectangular pulse controls the rise time [\subfig{1}{d}]. At long (short) rise times, pulses appear Gaussian (almost rectangular).
 Standard spin to charge conversion techniques are used in the region S(2,0) during spin projection measurements to obtain the singlet
occupation probability $P_{S}$\cite{Ono2002}. Applying a detuning pulse of duration $\tau$ will result in a phase accumulation between
the quantum state components. This phase is related to both the accumulation time and the detuning voltage.\cite{Gaudreau2011,Ribeiro2009,Ribeiro2010,Sarkka2011,Petta2010}

The dependence of the oscillations on rise time is shown in the left panels of \subfig{2}{a,b} for a large enough detuning pulse to
 allow mixing with the S/$\Tzero$ states (the graphs in this letter use the pulse detuning $dV_2$ component defined with respect to the
observed charge transfer line). For optimum observation of the  S/$\Tplus$ oscillations a rise time of a few ns is usually required as shown in the
 left hand panel of \subfig{2}{a}.  This is due to a competition between the  Landau-Zenner tunneling probability during the passage through the anticrossing and the restriction
imposed by the coherence  time. \cite{Gaudreau2011}  The S/$\Tplus$  fringes have a negative slope in the $\tau$-$\Vtwo$ plane. However, if the rise time is shortened to less than 1 ns an
interplay between the two qubit species becomes manifest. The S/$\Tplus$ oscillations weaken (as the Landau-Zener
tunnelling probability approaches unity \cite{Shevchenko2010, Zener1932}) and S/$\Tzero$ oscillations with a positive
gradient appear across the S/$\Tplus$ oscillations [see left panel \subfig{2}{b}]. The calculated results agree
well with experimental data as shown in the right panels of \subfig{2}{a,b}. The opposite
slopes for the two qubits occur because as a function of detuning the level spacing decreases in one qubit while increasing in
the other (see \subfig{1}{c}).

\begin{figure}
\includegraphics[width=0.9\columnwidth]{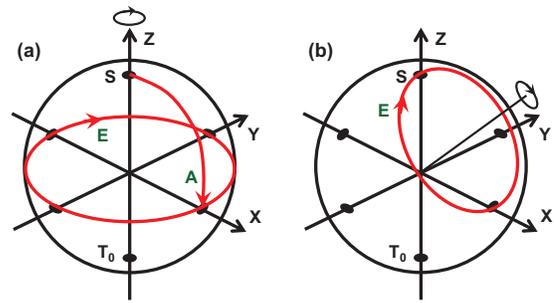}\vskip 0cm
\caption{(a): Schematic of S/$\Tzero$ Bloch sphere showing state vector motion corresponding to steps A and E of the ''spin
swap" pulse of \subfig{3}{c}. Ideally optimized axis of rotation during the exchange step E is around the Z-axis. (b): Schematic state
vector motion corresponding to the ''fast" square pulse of \subfig{1}{d}. Axis of rotation is lowered towards the equator.}
\label{fig:4} \end{figure}

On varying the interdot coupling we find the contrast visibility of the S/$\Tzero$ fringes improves substantially. We speculate that this
results from less sharp detuning dependence of the S/T$_0$ splitting and hence a lower sensitivity to charge noise effects.
 We therefore switch to a voltage configuration in which the conditions for the  two dot approximation  are
still upheld but where the interdot coupling is stronger, in this case  61 $\mu$eV compared to
17 $\mu$eV for \subfig{2}{a,b}. Due to this larger coupling the applied magnetic field range is also larger in this
regime and fields up to 1T may be used (in what follows we use B = 0.2~T). In this regime more oscillations are visible
and the interplay clearer. We first demonstrate the S/$\Tzero$ qubit alone. To achieve this  we perform a variation of
the "spin-swap" scheme \cite{Petta2005} using a pulse, illustrated in \subfig{3}{c}, consisting of a fast 12~mV segment
to cross the S/$\Tplus$ from the initial detuning $dV_2$ to $\epsilon$=$dV_2$+12~mV followed by a 100~ns adiabatic
elliptical pulse to $dV_2$+25~mV  in the S/$\Tzero$ interaction region (region A in \subfig{3}{d}) and then a rapid
step backwards to $dV_2$+17~mV (region E in \subfig{3}{d}) where the finite exchange splitting causes rapid oscillations
during a time $\tau$. The pulse then follows the reverse path back to the initial detuning for spin to charge
readout. The adiabatic step in this scheme rotates the state vector on the S/$\Tzero$ Bloch sphere down to the equator
which maximizes the effect of exchange rotation (see \subfig{4}{a}). The resulting fringes are shown in \subfig{3}{a}
to be compared to the calculation in \subfig{3}{b}. The internal structure of the fringes in theory and experiment (see
circled region in \subfig{3}{a}) is related to slight non-adiabaticity in the elliptical part of the pulse. We observe
S/$\Tzero$ oscillations  persisting to  25~ns in \subfig{3}{a}.

\begin{figure}
\includegraphics[width=1.0\columnwidth]{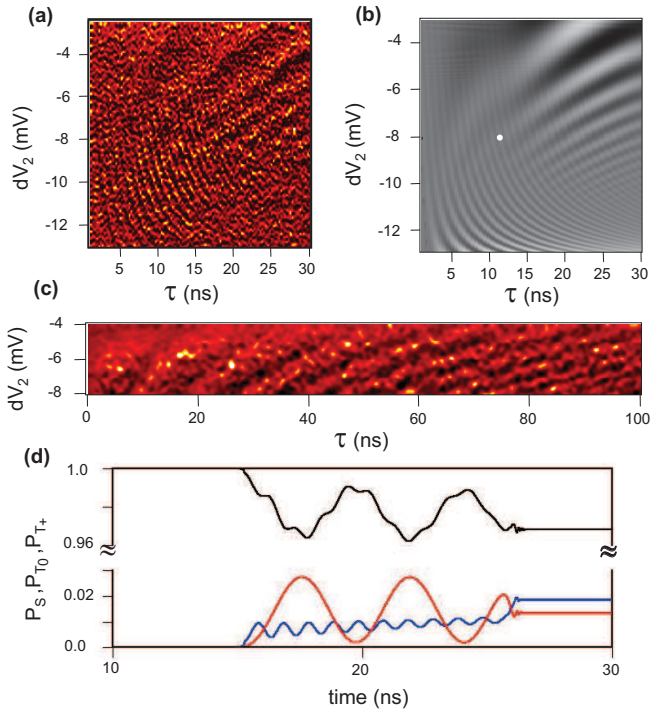}\vskip 0.0cm
\caption{(a): Experimental map in the  $\tau-\Vtwo$~plane from the numerical derivative of the QPC conductance with respect to detuning component $\Vtwo$ for a pulse rise time of 1.0 ns. The pulse involves ($\delta V_1$,
$\delta V_2$)=(-16.2,20.0)~mV to traverse the charge transfer line between (2,0) and (1,1) and repeats every $T_m$=2~$\mu$s, B=0.2~T.  Black is low, orange is medium, and yellow is high transconductance. (b): Calculated derivative with respect to detuning of $P_{S}$ in the $\tau-\Vtwo$ plane  for the same rise time as in (a).  Black (white) is low (high). The origin for the  detuning voltage  is the charge transfer line. (c): Narrow detuning region with rectangular pulse, ($\delta V_1$, $\delta V_2$)= (-19.5,25.0)~mV,    rise time $\approx$ 0.25 ns.  (d): Time response of $P_{S}$ (black), $P_{T_{0}}$ (red) and $P_{T_{+}}$ (blue) to single pulse at point on \subfig{5}{b} indicated by white dot.} \label{fig:5} \end{figure}

Having observed S/$\Tzero$ oscillations in the stronger coupling regime we return to a fast pulse scheme as in \subfig{1}{d} to observe
interactions between all 3 states. The experimental results in this regime are shown in  \subfig{5}{a} for a 1~ns pulse rise time.
At large negative detuning over twenty S/$\Tplus$ coherent oscillations are seen.

At less negative detuning strong S/$\Tzero$ oscillations are seen without recourse to the initial adiabatic initialization
step introduced in Ref.\cite{Petta2005} to rotate the state vector to the equator on the Bloch sphere as illustrated in \subfig{4}{a}.
 We note, however, that our S/$\Tzero$ fringes are essentially the same as those observed in the adiabatic initialization scheme although
the period is larger because the detuning range extends to regions of smaller splitting between S and T$_0$.  Such a pulse shape produces
strong oscillations in the weak exchange interaction regime close to the S/$\Tzero$ asymptotic region where the axis of qubit rotation on
the S/$\Tzero$ Bloch sphere is not purely around
the Z axis but is tilted downwards towards the equator producing enhanced visibility (see  \subfig{4}{b}). At intermediate  detunings we
observe the regime where the two qubits coexist and qubit interplay is clearly visible. The $\Tts$ for the S/$\Tplus$ interaction in the
theoretical calculation is 10 ns which is similar to values found in previous work on double \cite{Petta2010} and triple dots \cite{Gaudreau2011}.
Interestingly the S/$\Tzero$ fringes in the region where the energy splitting is small appear to persist to much longer times in contrast to a value of 10~ns quoted in \cite{Petta2005} and in \subfig{5}{b} we use an infinite $\Tts$ for S/$\Tzero$. In fact, as illustrated in \subfig{5}{c},
we frequently  see oscillations for pulse durations exceeding 100~ns when the rise time is very short. We find $\Tts$ is shorter for smaller
detunings consistent with charge noise limiting an enhanced coherence time. It has been suggested that such an enhancement
could originate either from non-zero exchange and/or an incidental narrowing of the nuclear environment \cite{Coish2005, Coish2009}. To see the detailed nature of the oscillations involving both S/$\Tzero$ and S/$\Tplus$ interactions we choose a
point on the calculated map \subfig{5}{b} indicated by the white dot and plot the time dependence of the state occupation
probabilities $P_{S}$, $P_{T_{0}}$ and $P_{T_{+}}$ as a function of time before, during and after the pulse. This theoretical analysis of the
quantum interference process finds the common component singlet state, S, is strongly affected by both qubit
modulations while the two triplet states fundamentally retain their individual oscillatory character.

In conclusion we have studied a scenario where two qubits, S/$\Tzero$ and S/$\Tplus$, sharing a common component state,
coexist  and interplay. The main feature in this evolution of a three state system is a  quantum interference
effect. The relative strength of each qubit contribution can be tuned with the rise time of the pulse responsible for
the quantum state preparation. Simulations provide good agreement with the experimental results.
The $T_{0}$ and $T_{+}$  components show the individual oscillations for the appropriate interaction while the S component shows
both.
 The S/$\Tzero$  qubit, triggered by the perpendicular component of the statistical nuclear field gradient persists an order of
magnitude longer than the S/$\Tplus$ qubit driven by the in-plane component.  This longer coherence time is consistent with a lower sensitivity  to charge noise.

We acknowledge discussions with Bill Coish, Michel Pioro-Ladri\`ere, Aash Clerk, Guy Austing and Roland Brunner, and O. Kodra
for programming. A.S.S. acknowledges funding from NSERC and CIFAR. G.G. acknowledges funding from the NRC-CNRS
collaboration.

\end{document}